\documentclass[aps,twocolumn,superscriptaddress,showpacs]{revtex4}
\usepackage{amsmath, amsfonts, amsthm}
\usepackage{graphicx}

\newcommand{\psibar}{\overline{\psi}}
\newcommand{\phibar}{\overline{\phi}}
\newcommand{\be}{\begin{eqnarray}}
\newcommand{\ee}{\end{eqnarray}}
\newcommand{\ea}{\end{array}}
\newcommand{\ba}{\begin{array}}
\newcommand{\tperp}{t_{\bot}}
\newcommand{\tpar}{t_{\|}}

\begin{document}
\graphicspath{{image/}{./}}
\title{Regular networks of Luttinger liquids}

\author{K. Kazymyrenko}
\author{B. Dou\c{c}ot}
\affiliation{Laboratoire de Physique Th\'{e}orique et Hautes
\'{E}nergies, CNRS UMR 7589,  \\
Universit\'{e} Paris VI and VII,
4 place Jussieu, 75252 Paris Cedex 05, France}

\date{\today}

\begin{abstract}
We consider arrays of Luttinger liquids, where each node
is described by a unitary scattering matrix.  
In the limit of small electron-electron interaction,
we study the evolution of these scattering matrices
as the high-energy single particle states are gradually
integrated out. Interestingly, we obtain the same renormalization
group equations as those derived by Lal, Rao, and Sen, for
a system composed of a single node coupled to several semi-infinite
1D wires. The main difference between the single node geometry
and a regular lattice is that in the latter case, the single
particle spectrum is organized into periodic energy bands, so
that the renormalization procedure has to stop when the last
totally occupied band has been eliminated.
We therefore predict a strongly renormalized Luttinger liquid
behavior for generic filling factors, which should exhibit
power-law suppression of the conductivity at low temperatures
$E_{F}/(k_{F}a) \ll k_{B}T \ll E_{F}$, where $a$ is the lattice 
spacing and $k_{F}a \gg 1$. Some fully insulating ground-states
are expected only for a discrete set of integer
filling factors for the electronic system. A detailed discussion
of the scattering matrix flow and its implication for the low
energy band structure is given on the example of a square lattice.

\end{abstract}

\pacs{To be determined}

\maketitle

\section{Introduction}
For the past two decades, transport properties of quantum wires have received
a lot of attention~\cite{Timp91,Washburn92}. Besides metallic systems, two dimensional
electron gases induced in GaAs/AlGaAs heterostructures have displayed a rich
variety of quantum effects such as Aharonov-Bohm resistance oscillations in
a ring geometry~\cite{Timp87,Pedersen00}, and persistent currents~\cite{Mailly93,Rabaud01}.
Since the electronic transport mean free path in such artificial nanostructures
can be as large as several micrometers, most scattering processes for electronic
quasiparticles occur at the nodes between several conducting wires. Their influence has
been extensively studied in the context of the breakdown of the Quantum Hall Effect
in narrow channels. Experiments have revealed that the Hall resistance measured in
a four probe geometry disappears at low magnetic fields~\cite{Roukes87,Ford88}.
Theoretical studies have emphasized the role of quantum mechanical resonances
in the scattering amplitudes of electrons at the junctions between the main channel
and voltage probes~\cite{Ravenhall89,Kirczenow89}. In experimental systems, confining
potentials remain smooth in the vicinity of such junctions, and this induces
a rather robust collimation mechanism for incoming electrons~\cite{Baranger89,Beenakker89}.
This semi-classical picture has been confirmed by spectacular experiments involving
junctions with various shapes~\cite{Ford89}. More recently, coherent Aharonov-Bohm
oscillations have been measured in ballistic arrays with the dice lattice 
geometry~\cite{Naud01}, in agreement with the predictions of simple models
for non-interacting electrons~\cite{Vidal98,Vidal00}.

In parallel to this mostly single electron physics, 
dramatic electron-electron interaction effects have
been demonstrated in transport measurements on various ballistic
conductors with very few transverse conduction channels.
For instance, tunneling into edges of a two-dimensional electronic
droplet in the Fractional Quantum Hall Effect (FQHE) regime
has shown current versus voltage curves with power law behavior~\cite{Grayson98}
in qualitative (though not really quantitative) agreement with theoretical
models based on the chiral Luttinger liquid picture~\cite{Wen90}.
Measurements of shot noise associated to tunneling processes from
one edge to another through a quantum point contact have also
provided a convincing demonstration of the presence of fractionally
charged quasiparticles in the FQHE phase~\cite{Kane94,Picciotto97,Saminadayar97}.
Another family of one-dimensional quantum conductors are carbon nanotubes.
In particular, single wall nanotubes have shown a strong reduction of
the single particle density of states at low energies~\cite{Bockrath99,Yao99},
compatible with the Luttinger liquid model~\cite{Egger97,Kane97}.

These two main lines of research just outlined can be naturally
combined and lead us to consider the subject of networks of
interconnected quantum wires, each of them being described as
a Luttinger liquid. As a first step in this direction, several systems
with nanotube crossings have been synthesized~\cite{Papadopoulos00,Kim01,
Terrones02,Gao03}. On the theoretical side, many studies of Luttinger
liquids crossing at one node are now available~\cite{Komnik98,Nayak99,Lal02,
Chen02,Chamon03}, including extensions to more complex geometries~\cite{Das03}.
In this paper, we consider a regular network of Luttinger liquids.
As already mentioned, the main source of electron scattering in ballistic
structures arises from the nodes of the network. For non-interacting
electrons, these nodes are simply described by a scattering 
matrix~\cite{Schult90,Deo94,Uryu96}, and the full band structure
(in the absence of disorder) can be retrieved from the knowledge of this
matrix. However, as first shown by Kane and Fisher for a single
impurity in a Luttinger liquid, interaction effects induce a variation
of the dressed scattering matrix as a function of the incoming electron 
energy~\cite{Kane92a,Kane92b}. One way to interpret this in physical terms is via 
the notion of Anderson's orthogonality catastrophe: in the limit of a tunnel barrier,
an electron jumping across the barrier leaves a dipolar charged excitation
which is very far from any eigenstate of the interacting system.
A rather complicated collective relaxation process follows any single
electron tunneling event. A remarkable prediction made in these works is
a dramatic qualitative difference between repulsive and attractive interactions.
In the former case, the effective impurity potential grows as the typical
energy becomes closer to the chemical potential. So a single impurity
is sufficient to disconnect completely an infinite Luttinger liquid at $T=0$
for repulsive interactions. Conversely, any static impurity becomes
transparent in the low energy limit in the case of attractive interactions. 

An appealing picture for these effects has been proposed by
Yue, Glazman and Matveev~\cite{Yue94}. They have shown that 
renormalization of the transmission amplitude may be attributed 
to scattering of an incoming electron on Friedel density oscillations
induced by the scatterer. From this picture, they have developed
an alternative renormalization approach, which is perturbative
in the electron-electron interaction, but non-perturbative in the strength
of the impurity potential. This framework has been used later in 
references~\cite{Lal02,Das03}, and we shall adopt a similar
procedure here. Note that a third type of renormalization scheme,
involving the full momentum dependence of the electronic self-energy,
has been implemented in a series of papers~\cite{Meden02a,Meden02b,Meden03}. 

The main novel feature in regular arrays in comparison to simpler
geometries as a few connected wires is the existence of commensurability
effects between the Fermi wave-length of electrons and the lattice
period. In a non-interacting electron picture, we expect an energy gap in the
spectrum when the average electron number in each unit cell of the lattice is
an integer. As we shall see later, the band structure for
a two-dimensional network yields a gapped excitation spectrum
whenever some integer numbers of bands are filled.
For interacting electrons, commensurability effects may also be understood
by considering the pattern of Friedel density oscillations. 
In a one-dimensional geometry,
these oscillations exhibit a dominant wave-vector equal to $2k_{F}$,
where $k_{F}$ is the Fermi wave-vector for a non-interacting one-dimensional
wire with the same electronic density. Let us denote by $a$ the distance between 
two nodes. Friedel oscillations originating from different nodes share
the same global phase if $2k_{F}a$ is an integer, which simply means that
the average number of electrons along a segment of length $a$ is integer.
Therefore, in the case of repulsive interactions, we expect an insulating
ground-state in the commensurate case, where the Kane-Fisher mechanism
will disconnect all the wires incoming at the same node. For incommensurate
fillings, we predict a strongly renormalized Fermi liquid, where the partially
filled band crossing the Fermi level becomes much less dispersive than for the
original non-interacting band structure. We suggest that these effects should be
in principle observable in networks of ballistic wires where the electronic density
could be controlled by an uniform gate potential. By changing the gate voltage, these
systems are expected to undergo a succession of metal-insulator transitions.
The difference between an interacting system and a non-interacting one
will be manifested by power-law dependences for the conductance as a function
of temperature at fixed bias voltage, or as a function of $V$ at fixed $T$~\cite{Kane92a,Kane92b},
provided both $k_{B}T$ and $eV$ remain higher than an energy scale
$\Delta$ which is the renormalized band splitting in the incommensurate case,
or the single particle gap in the commensurate one.

This paper is organized as follows. In section~\ref{1Dlat}, we consider a simpler problem,
namely a one-dimensional chain of regularly spaced impurities.
We set up a renormalization group method for weakly interacting electrons
which is closely related to those developed in Refs~\cite{Yue94,Lal02}, but where
the periodicity of the system is explicitly taken into account.
Section~\ref{generallat} generalizes this approach to any lattice composed
of links of the same length, assumed to be large compared to the Fermi 
wave-length. We show explicitly that the scattering matrices at each node
of such lattices is renormalized exactly in the same way as 
for a single node connecting semi-infinite wires~\cite{Yue94,Lal02}.
This is the central result of the present work. 
As an illustration with possible experimental relevance, 
section~\ref{2Dlat} considers a two-dimensional square lattice of Luttinger liquids. 
We show that although the evolution of the scattering matrix
of the nodes as the typical energy scale is reduced yields
a rather trivial low-energy fixed point where all the links become
disconnected, some interesting qualitative changes in the quasiparticle
band structure take place along this renormalization group flow.

\section{One dimensional wire with a periodic impurity potential}
\label{1Dlat}

The goal of this section is to adapt the simple renormalization group
procedure initiated in Refs.~\cite{Yue94,Lal02} to the case of
a periodic potential. The main idea developed in these works is to
dress the bare scattering amplitude by a correction due to the interaction
of an incoming electron with the Friedel density oscillation induced by the
impurity. This approach treats the electron-electron interaction as the
perturbation. Because the continuous spectrum of particle-hole excitations
in the metallic wire exhibits a finite density of states down to  
arbitrary low energy, the first order correction to the scattering amplitude
diverges as \mbox{$\ln(|k-k_{F}|d)$}, 
where $k$ is the incoming electron's wave-vector,
and $d$ is the spacial range of the bare impurity potential. This type of
infra-red divergence is very similar to those encountered in the Kondo problem,
and Yue et al. proposed to treat them with a renormalization group method
inspired by Anderson's ``{\em poor man's scaling}'' approach~\cite{Anderson70}.
The idea is to integrate out gradually single particle-hole excitations
which participate in the Friedel oscillation, starting from those furthest
from the Fermi level. As the electron bandwidth $D$ is continuously reduced,
the bare impurity potential is renormalized so that the low energy physical
properties of the system are kept unchanged. The renormalization procedure
stops at a low energy scale with is the larger scale among the thermal
broadening $k_{B}T$, the bias voltage $eV$, or the incoming electron energy
$\hbar|k-k_{F}|v_{F}$.

As already stated in the Introduction, the presence of an array of
scattering centers (such as nodes in a wire network) brings 
qualitatively new features. In the low energy regime, Friedel oscillations
originating from different centers are expected to interfere, so
we cannot follow the renormalization flow obtained in Refs~\cite{Yue94,Lal02}
for a single scatterer down to arbitrary low energies.
Furthermore, commensuration effects between the average electronic density
and the superlattice structure play a crucial role. By contrast to the
single impurity case, we expect an insulating ground-state only for
an integer average filling of each supercell. 
For incommensurate filling factors, we expect
a cross-over from the one-dimensional behavior following Kane and Fisher's
predictions at high energy, towards a strongly renormalized coherent conductor
at low energy with finite conductance. 
In a periodic system, the natural way to implement this ``{\em poor man's scaling}''  
approach is to integrate out energy bands one after the other, starting from
those most remote from the Fermi level. In the incommensurate case, the
last band, which crosses the Fermi level is partially filled, so it is natural
to stop the procedure after the last fully occupied band has been integrated out.
In any renormalization method, we have to decide which low-energy quantities
will be required to remain constant as high energy modes are eliminated. In the
presence of a periodic potential, it is natural to prescribe that single
quasiparticle energies should not change under the renormalization group flow (RGF).

 \subsection{Band structure for a periodic array of point scatterers}

Let us first consider a non-interacting problem along an infinite one-dimensional
wire with a periodic potential. The corresponding Hamiltonian is
\begin{equation}
\hat H=\frac{\hat p^2}{2m}+\sum_{n=0}^{N-1}V(x-na)
\end{equation}
where $a$ denotes the spacial period of the potential, namely the
distance between two succesive impurities. 
$V(x)$ is a localized potential, so for instance we impose that 
$V(x)=0$ when $|x|$ is larger than a range $d$, $d \ll a$.
The effect of each scatterer is described by a scattering matrix $\hat{S}$.
Suppose first we have only one of them, centered at the origin $x=0$.
Let us consider scattering states with the energy $E_{0}(k)=\hbar^{2}k^{2}/(2m)$,
$k$ being positive. Away from the impurity, that is for $|x|>d$, we may
represent the corresponding wave-function as a superposition of plane-waves:
\begin{equation}
\psi(x)=\left\{\begin{array}{ll}
Ae^{ikx}+Be^{-ikx} & \mbox{for $x<-d$} \\ 
A'e^{ikx}+B'e^{-ikx} & \mbox{for $x>d$}
\end{array}\right.
\end{equation}
\begin{figure}[bt]
\includegraphics[width=2.7in]{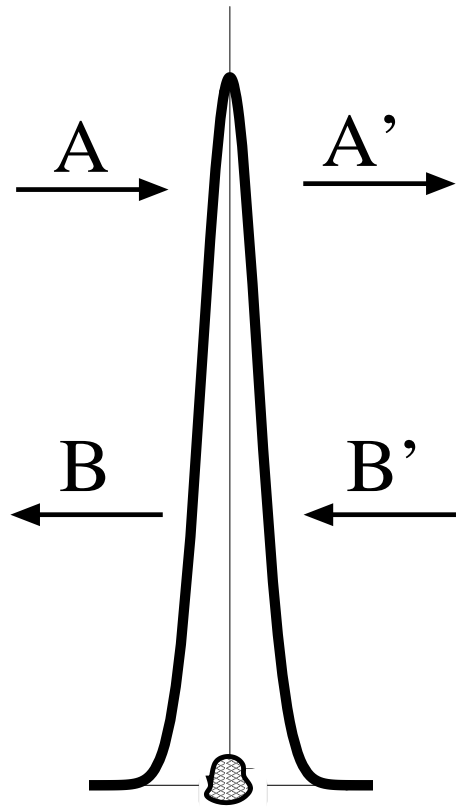}
\caption
{Localized impurity potential may be represented by a $\hat S$-matrix, 
that connects amplitudes of incoming ($A,B^\prime$) and outgoing 
($A^\prime,B$) plane waves outside the impurity.}
\label{ABBA_xfig}
\end{figure}

Since Schr\"odinger's equation is linear and of second order,
we may express the outgoing amplitudes $A'$ and $B$ linearly
as a function of the incoming ones $A$ and $B'$
\begin{equation}\label{s-matrix}
\left(\ba{c}A' \\ B \ea \right)= 
\left( \ba{cc} t & r' \\ r & t'\ea \right) 
\left(\ba{c} A \\ B' \ea \right) 
\equiv \hat{S} \left( \ba{c} A \\ B' \ea \right)
\end{equation}
where $\{r,t,r',t'\}$ are two pairs of reflection and transmission coefficients for left
and right sides of the node. In principle, these four coefficients do depend on the
energy of the particle or equivalently on its wave-vector $k$. In this paper, we
shall neglect this variation, since the dominant contribution processes involve virtual
excitation of particle-hole pairs in the vicinity of the Fermi level. A more complete
approach would consider the Taylor expansion of $\hat{S}$ in powers of $k-k_{F}$, but
all terms beyond the 0'th order one are irrelevant according to the classification
of perturbations around a non-interacting one-dimensional fermion system.
At least for not too large interactions, they are not supposed to change the
qualitative picture of the system behavior.
As usual, this $\hat{S}$ matrix is unitary. Assuming time reversal invariance
of the Hamiltonian implies $t=t'$ and if $V(x)$ is an even function of $x$,
we have also $r=r'$. In this case, we may parametrize $\hat{S}$ by two angles
($0\leq \phi \leq \pi/2,\; 0\leq \psi <2\pi$)
\begin{equation}\label{s-paramet}
\hat{S}=e^{i\psi}\left( \ba{cc}
  \cos\phi & \pm i \sin\phi\\
  \pm i\sin\phi & \cos\phi
\ea \right)
\end{equation}

For a periodic array of identical scatterers, eigenstates may be obtained
as Bloch functions, namely we may impose the condition
\[\psi(x+a)=e^{ik'a}\psi(x),\] 
where $k'$ is chosen in the first Brillouin zone $\left[-\pi/a,\pi/a\right]$.
On each $x$-interval $\left[na+d,(n+1)a-d\right]$, we write the eigenstate
with energy $E_{0}(k)$ as
\[\psi(x)=A_{n}e^{ikx}+B_{n}e^{-ikx}.\]
The above periodicity condition implies
\begin{eqnarray*} 
A_n & = & e^{i(k'-k)an}A_0 \\ 
B_n & = & e^{i(k'+k)an}B_0 
\end{eqnarray*}
Eq.~(\ref{s-matrix}) can now be written for each impurity site, which gives:
\begin{equation}\label{s-matrixn}
\left(\ba{c}A_{n+1}e^{ika(n+1)} \\ B_{n}e^{-ika(n+1)} \ea \right)= 
\left( \ba{cc} t & r' \\ r & t'\ea \right) 
\left(\ba{c} A_{n}e^{ika(n+1)} \\ B_{n+1}e^{-ika(n+1)} \ea \right) 
\end{equation}
Replacing $A_{n}$ and $B_{n}$ by their expressions in terms of
$A_{0}$ and $B_{0}$, we get the following secular equation
\begin{equation}\label{det}
\left| \ba{cc}
  te^{i(k-k')a}-1 & r'\\
  re^{i2ka} & t'e^{i(k+k')a}-1
\ea \right|=0
\end{equation}
which determines the dispersion relation implicitly via
$k$, the energy being $E_{0}(k)$, the lattice momentum $k'$ 
behaving as an external parameter.  
Using a normalization condition on the
wave-function, we could get $(A_0,B_0)$ as functions of
$(k',\hat S)$.

In the particular case of spacially even and time-reversal invariant
potentials, we may use the above parametrization for $\hat{S}$ 
in Eqn.(~\ref{det}), which yields
\begin{equation}\label{band1D}
\cos(ka+\psi)=\cos\phi\cos(k'a)
\end{equation}
For a given value of the lattice momentum $k'$, the possible values
of $ka$ appear in two equally spaced families, with a period $2\pi$
for each of them. The allowed values of $ka+\psi$
belong to the intervals $[-\pi+\phi+2\pi n,-\phi+2\pi n]$ and $[\phi+2\pi n,\pi-\phi+2\pi n]$,
where $n$ is integer. We recall that $k$ should be {\em positive}, in order
not to count each eigenstate twice. 
The values of $ka+\psi$ lying in intervals $[-\phi+2\pi n,\phi+2\pi n]$, 
and $[\pi-\phi+2\pi n,\pi+\phi+2\pi n]$  correspond then to energy gaps.
These gaps are of course larger when the reflexion coefficient is larger.
\begin{figure}[bt]
\includegraphics[width=2.7in]{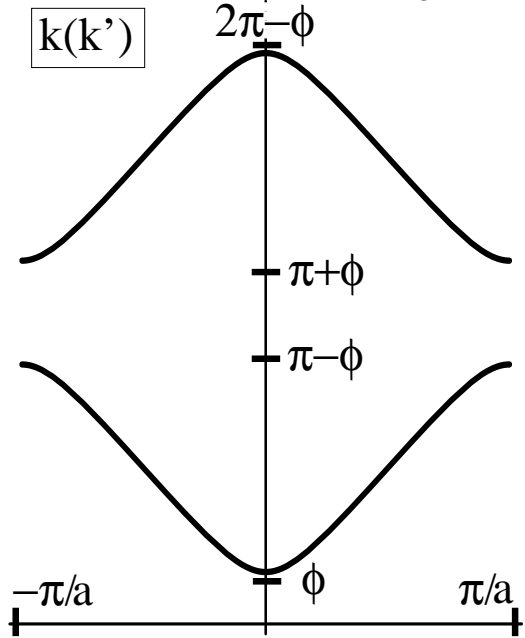}
\caption
{Band structure for 1D wire of noninteracting electrons with periodic impurities. BS is $2\pi$ periodic in $k$. A set of gaps $(-\phi+2\pi n,\phi+2\pi n) \text{ and }(\pi-\phi+2\pi n,\pi+\phi+2\pi n) $ is present for any value of $\hat S$-matrix.} 
\label{band1Dexact_xfig}
\end{figure}

 \subsection{Switching on electron-electron interactions}

In a Luttinger liquid, the effective interaction becomes
non-local in the low-energy limit. To show this, it is
convenient to decompose the electron creation operators
$\Psi^{+}_{\sigma}(x)$ (where $\sigma \in \{\uparrow,\downarrow\}$
denotes the spin component along a fixed direction) into a
right moving part $\Psi^{+}_{R\sigma}(x)$ and a left-moving part
$\Psi^{+}_{L\sigma}(x)$, where $\Psi^{+}_{R\sigma}$ (resp. 
$\Psi^{+}_{L\sigma}$) involves the Fourier modes $k$ close to
$k_{F}$ (resp. $-k_{F}$). With this decomposition, the local
electron density $\rho(x)$ is written as follows
\begin{widetext}
\begin{eqnarray}
\rho(x)=\sum_{\sigma=\uparrow,\downarrow}\Psi^{+}_{\sigma}(x)\Psi_{\sigma}(x)
  =  \sum_{\sigma=\uparrow,\downarrow}
\left(\Psi^{+}_{R\sigma}(x)\Psi_{R\sigma}(x)+\Psi^{+}_{L\sigma}(x)\Psi_{L\sigma}(x)\right)+
\sum_{\sigma=\uparrow,\downarrow}\left(\Psi^{+}_{R\sigma}(x)\Psi_{L\sigma}(x)
+\Psi^{+}_{L\sigma}(x)\Psi_{R\sigma}(x)\right) 
\end{eqnarray}
The first two terms are smooth fields, meaning that their Fourier
transforms involve only small wave-vectors compared to $k_{F}$.
But the last two terms are centered around the wave-vectors
$\pm 2k_{F}$ so they are rapidly oscillating. 
\end{widetext}
For a spin-rotation
invariant Hamiltonian, the effective low energy description of a 
Luttinger liquid involves three independent parameters:
the velocities $v_{c}$ and $v_{s}$ of collective charge and spin
excitations, and a dimensionless constant $K$ which depends
on the strength of electron-electron interactions and controls
the exponents entering the correlation functions. Since transport
properties are mostly affected by the value of $K$~\cite{Kane92a},
we shall not consider here the renormalizations of $v_{c}$ and $v_{s}$
away from their common value $v_{F}$ for a non-interacting system.
Therefore, it is sufficient to consider the following interaction
\begin{equation}\label{HintLut1}
H_{\mathrm{int}}=\frac{U_{0}}{2}\int\limits_{-L/2}^{L/2} dx \rho_{0}(x)^{2} 
\end{equation}
where $\rho_{0}(x)$ is the long wave-length part of the total density:
\begin{equation}
\rho_{0}(x)=
\sum_{\sigma=\uparrow,\downarrow}\left(\Psi^{+}_{R\sigma}(x)\Psi_{R\sigma}(x)
+\Psi^{+}_{L\sigma}(x)\Psi_{L\sigma}(x)\right) \nonumber
\end{equation}
Here $L$ denotes the total length of the system. Later, we shall
assume periodic boundary conditions, and that $L$ encloses an integer
number $N$ of periodic cells, so $L=Na$.
With this choice of interaction, we have 
\begin{eqnarray*}
v_{c} & = & v_{F}\left(1+\frac{2U_{0}}{\pi \hbar v_{F}}\right)^{1/2} \\
v_{s} & = &v_{F} \\
K & = & \left(1+\frac{2U_{0}}{\pi \hbar v_{F}}\right)^{-1/2}
\end{eqnarray*}
So $K=1$ for a non-interacting system, $K>1$ for attractive interactions,
and $K<1$ for repulsive interactions. For our purpose, it is convenient
to view this effective interaction as deriving from a {\em non-local}
potential $U(x-y)$ such that its Fourier transform $\tilde{U}(k)$
vanishes outside a finite window centered around $k=0$ and whose
width is smaller than $2k_{F}$. The interaction strength $U_{0}$
is defined as $\tilde{U}(k=0)$. With this notation, we have 
\begin{equation}\label{HintLut2}
H_{\mathrm{int}}=\frac{1}{2}\int\limits_{-L/2}^{L/2} dx \int\limits_{-L/2}^{L/2} dy
\rho_{0}(x)U(x-y) \rho_{0}(y)
\end{equation}

In this section, we are considering the combined effect of
impurity scattering and interactions. Renormalizations
of the effective scattering matrix $\tilde{S}$ are naturally
detected via the electron self-energy $\Sigma(k,k',\omega)$.
But since our system exhibits only a discrete translation symmetry,
we may only conclude that $k'-k$ should be an integer multiple
of the basic reciprocal lattice vector $\frac{2\pi}{a}$.
This self-energy is then a relatively complicated object.
More information on its real-space structure for a single impurity
may be found in~\cite{Meden02a,Meden02b}.
To analyze it in a simple way we shall compute the first order 
correction $E_{1}(k)$ with respect to $U_{0}$ to the single
electron energy $E_{0}(k)=\hbar^{2}k^{2}/(2m)$. Here $k$
stands for a single particle level, close to the Fermi energy,
and labelled by the combination of a Bloch quasi-momentum $k'$
and a band index. This correction $E_{1}(k)$ 
is given by the sum of a Hartree term and of an exchange term.
In the case of an unpolarised electron system, we have
\begin{widetext}
\begin{equation}\label{E1simple}
E_{1}(k)=\int\limits_{-L/2}^{L/2} dx \int\limits_{-L/2}^{L/2}dy\sum_{q<k_F}
\psi_k^*(x)\psi_q^*(y) U(x-y)[2\psi_q(y)\psi_k(x)-\psi_q(x)\psi_k(y)]
\end{equation}
\end{widetext}
For local potentials, the Hartree and the exchange contributions
cancel each other, when the spins of the two electrons involved
are parallel. But as we have recalled before, our two-body effective potential
is in fact non-local, so we have to analyze both terms in more detail.
Our expression for $E_{1}(k)$ involves integrals of the form:
\[
I_{U}(f,g)=\int\limits_{-L/2}^{L/2} dx \int\limits_{-L/2}^{L/2}dy
f^{*}(x)U(x-y)g(y)
\]
where $f(x)$ and $g(y)$ are Bloch functions satisfying
\[
\frac{f(x+a)}{f(x)}=\frac{g(y+a)}{g(y)}=e^{i\theta},\qquad 0<\theta\le2\pi
\]
Since $U$ is short-ranged in space (though it is {\em not}
a delta function), we may take safely the thermodynamic limit
$L\rightarrow \infty$. Writing $f(x)=\sum_{n}\tilde{f}_{n}e^{i((2\pi n+\theta)x/a)}$
and an analogous series for $g(y)$, we obtain:
\begin{equation}\label{filter}
I_{U}(f,g)=L\sum_{n}\tilde{f}^{*}_{n}\tilde{U}(\frac{2\pi n+\theta}{a})\tilde{g}_{n}
\simeq LU_{0}\sum_{|n| \ll k_{F}a}\tilde{f}^{*}_{n}\tilde{g}_{n}
\end{equation}

Let us first consider the Hartree term.
Because we chose single particle eigenstates of the Bloch
form, the corresponding local particle density is periodic with
period $a$.
A little elementary algebra shows that
\begin{equation}
\label{density}
|\psi_{k}(x)|^{2}=\frac{1}{Na}\frac{\sin(ka+\psi)\pm\sin(\phi)\cos(2k(x-\frac{a}{2}))}
{\sin(ka+\psi)\pm\sin(\phi)\sin(ka)/(ka)}
\end{equation}
for $0<x<a$.
As expected, the amplitude of the local density oscillation is
stronger when the bare reflexion coefficient is larger, or equivalently
when $|\sin(\phi)|$ is larger.
The $n^{\mathrm{th}}$ Fourier amplitude of this local density is equal
(for $n \neq 0$) to:
\[
\frac{\mathcal{A}_{k}}{2Na}\left(\frac{\sin(ka-\pi n)}{ka-\pi n}+
\frac{\sin(ka+\pi n)}{ka+\pi n}\right)
\]
where the numerical coefficient $\mathcal{A}_{k}$ is close to unity.
As shown in Eq.~(\ref{filter}) above, we are interested in the case
where $|n| \ll k_{F}a$, and since $k$ is close to $k_{F}$, this amplitude
is small by a factor $1/(k_{F}a)$. A similar conclusion holds for the 
Fourier amplitudes of $|\Psi_{q}(x)|^{2}$ if we assume that the
most important effects come from filled states where $q$ is close to $k_{F}$.
Therefore, we do not expect strong renormalizations coming from the Hartree term. 

Let us now turn to the exchange term. The product $\Psi^{*}_{q}(y)\Psi_{k}(y)$
is the sum of four oscillating terms proportional to $e^{\pm i(k-q)y}$ and
$e^{\pm i(k+q)y}$. The last two terms are fast oscillations which will be filtered out
by the non-local potential, as in Eq.(\ref{filter}). Keeping only the first two
oscillations, we can cast the exchange contribution to $E_{1}(k)$ as follows:
\begin{widetext}
\begin{equation}\label{fullE1}
E_1(k)=c\frac{U_{0}}{a} \left(k'_F+N_F\pi\right)-\frac{U_{0}}{2\pi a}
\sin^{2}\phi \int  \frac{dq'}{\sin(k+\psi)\sin(q(q')+\psi)}\frac{\sin(q(q')-k)}{q(q')-k}
\end{equation}
In this equation, we have replaced combinations 
such as $ka$, $qa$, by new dimensionless variables $k$, $q$.
The integral symbol stands for a summation over all the $N_{F}$ completely filled
bands, including possibly a last partially filled band with a
dimensionless momentum $k'_{F}$ such that $0\leq k'_F<\pi$.
For each completely filled band, the integration variable $q'$ runs
from $0$ to $\pi$, and $q$ in Eq.~(\ref{fullE1})
is a function of the lattice momentum $q'$ solution of the
dispersion relation~(\ref{band1D}). 
For the last partially filled band (incommensurate case),
the $q'$ integral runs from $0$ to $k_{F}'$.
As already mentioned, we have assumed that
parameters ($\psi,\phi$) are not depending on the incoming energy.
Note that contributions from the Hartree term will modify
only the numerical coefficient $c$ whose precise value is not
important here.

 \subsection{Renormalization approach}

Let us introduce the notation $\Lambda_{0}=\pi N_{F}$, which
plays the role of a large momentum cut-off.
As in all schemes inspired by Anderson's {\em ``poor man's scaling''},
we shall assume it is possible to construct a sequence of models
where filled bands are eliminated one after the other, starting from
the most remote from the Fermi level. When the first $n$ bands have
been eliminated, the new value of $\Lambda$  is set equal
to $\Lambda_{0}-\pi n$. 
At each step, we require that the quasiparticle energy 
$(E_{tot}(k)=E_0(k)+E_1(k)\equiv E_0(k)+U_{0}\mathcal E_1(k))$
for $k$ close to the Fermi wave-vector should remain unchanged.
To compensate for the reduction of the cut-off from $\Lambda_{0}$
to $\Lambda$, we have to adjust $\hat{S}$-matrix parameters 
$\{\psi,\phi\}$ so they become functions of running cutoff $\Lambda$.
This is expressed by the following prescription
\begin{eqnarray} \label{Etotal}
E_{tot}(\psi_0,\phi_0,k)=E_0(\psi(\Lambda),\phi(\Lambda),k)+U_{0}\mathcal
E _1(\psi(\Lambda),\phi(\Lambda),k,\Lambda)
\end{eqnarray}
Since for $U_{0}=0$, this condition implies 
$E_{0}(\psi_0,\phi_0,k)=E_{0}(\psi(\Lambda),\phi(\Lambda),k)$,
we see that in this case $(\psi_0,\phi_0)=(\psi(\Lambda),\phi(\Lambda))$
for any $\Lambda$, so we may write the following Taylor series
\begin{eqnarray*}
\psi(\Lambda,U_0,\psi_0,\phi_0,\Lambda_0) & \equiv & \psi_0
+U_{0}\psibar(\Lambda,\psi_0,\phi_0,\Lambda_0)+O(U_0^2)\\
\phi(\Lambda,U_0,\psi_0,\phi_0,\Lambda_0) & \equiv & \phi_0
+U_{0}\phibar(\Lambda,\psi_0,\phi_0,\Lambda_0)+O(U_0^2)
\end{eqnarray*}
We now try to keep band structure (\ref{Etotal}) unchanged for any $k$
\begin{eqnarray} \label{renorm1} E_0(\psi_0,\phi_0,k)+U_{0}\mathcal E
_1(\psi_0,\phi_0,k,\Lambda_0)=
E_0(\psi_0+U_{0}\psibar(\Lambda),\phi_0+U_{0}\phibar(\Lambda),k)+U_{0}\mathcal E
_1(\psi_0+U_{0}\psibar(\Lambda),\phi_0+U_{0}\phibar(\Lambda),k,\Lambda)& \nonumber
\end{eqnarray}
Keeping the first order terms in $U_{0}$ gives
\begin{eqnarray} \label{factor} \frac{\partial E_0}{\partial
\psi}\Big|_{\psi_0}\psibar(\Lambda)+ \frac{\partial E_0}{\partial
\phi}\Big|_{\phi_0}\phibar(\Lambda)=\mathcal E_1(\psi_0,\phi_0,k,\Lambda_0)-\mathcal
E_1(\psi_0,\phi_0,k,\Lambda)
\end{eqnarray}
\end{widetext}
This is a non trivial constraint, since $\frac{\partial E_0}{\partial\psi}|_{\psi_0}$ 
and $\frac{\partial E_0}{\partial \phi}|_{\phi_0}$ depend on $k$
but do not depend on $\Lambda$. 
On the contrary, $\psibar$ and $\phibar$ depend on $\Lambda$
but not on $k$.
The possibility to enforce this requirement is not obvious
a priori, and when it occurs, we may call our model {\em renormalizable}
(at least to this lowest order).

Let us now evaluate the right-hand side of this equation.
Suppose we integrate out just one band, then $\Lambda_{0}-\Lambda=\pi$,
which is assumed to be much smaller than $\Lambda$. 
While computing  $\mathcal E_1(\psi_0,\phi_0,k,\Lambda_0)-\mathcal
E_1(\psi_0,\phi_0,k,\Lambda)$ in Eq.~(\ref{fullE1}),
the integral involves only one band
far from the Fermi level. Therefore, we may further approximate
$q(q')-k$ by $-\Lambda$. This yields
\begin{eqnarray}\label{varE11D}
\mathcal{E}_1(\psi_0,\phi_0,k,\Lambda_0)-\mathcal{E}_1(\psi_0,\phi_0,k,\Lambda)
\simeq \frac{c(\Lambda_{0}-\Lambda)}{a}+\nonumber \\
 \mbox{}+\frac{1}{\Lambda}\frac{1}{2\pi a}\frac{\sin^{2}\phi_{0}}{\sin(k+\psi_{0})}
\int_{0}^{\pi}dq'\frac{\sin(q(q')-k)}{\sin(q(q')+\psi_{0})}
\end{eqnarray}
From (\ref{band1D}) the derivatives involved in the left hand-side of Eq.~(\ref{factor}) are:
\begin{eqnarray*} 
\frac{\partial E_0}{\partial\psi}\Big|_{\phi,k=\mathrm{const}}
&\approx&-\frac{\hbar v_F}{a} \\
\frac{\partial E_0}{\partial \phi}\Big|_{\psi,k=\mathrm{const}}
&\approx&\frac{\hbar v_F}{a} \tan\phi\cot(k+\psi)
\end{eqnarray*}
We have linearized the bare dispersion relation:
$E_0(k)=\frac{\hbar^2k^2}{2ma^2}\approx const+\frac{\hbar v_F}{a}k$.
The notation ``$k=\mathrm{const}$'' means more precisely that the Bloch
crystal momentum $k'$ and the band index have to be maintained constant
while varying $\phi$ or $\psi$.
Introducing these expressions for the derivatives and the result~(\ref{varE11D})
in Eq.~(\ref{factor}) shows that indeed the $k$ dependences on both sides
can be made to match, which expresses the renormalizability of our model to 
first order in interaction strength. This fixes the form of the functions
$\psibar(\Lambda)$ and $\phibar(\Lambda)$:
\begin{eqnarray*}
\psibar(\Lambda,\psi_0,\Lambda_0) & = & \frac{1}{\hbar v_F}\big(c
(\Lambda-\Lambda_0)-\\&-&\frac{\sin^2\phi_0}{2\pi}\ln\frac{\Lambda}{\Lambda_0}
\int\limits_{0}^{\pi}dq'\cot(q(q')+\phi_0+\psi_{0})\big)\\
\phibar(\Lambda,\phi_0,\Lambda_0) & = & -\frac{1}{4\pi\hbar v_F}\sin
(2\phi_0)\ln\frac{\Lambda}{\Lambda_0}
 \end{eqnarray*}
Finally we construct the RGF equation:
\begin{eqnarray} 
&&\frac{\partial\psi}{\partial \Lambda}  =  
\frac{U_{0}}{\pi\hbar v_F}\big(c+\frac{\sin^2\phi}{2\pi}
\frac{1}{\Lambda}\int\limits_{0}^{\pi}dq'\cot(q(q')+\psi)\big)\nonumber\\
&&\frac{\partial\phi}{\partial\ln\Lambda} =  -\frac{U_{0}}{4\pi\hbar v_F}\sin 2\phi 
\end{eqnarray}
We see from Eq.~(\ref{s-paramet}) that the parameter $\psi$ is a global phase
in the scattering matrix, which does not affect any physical property of the system
besides an overall shift of the single particle spectrum.
In particular, it does not generate any density oscillation.
Moreover the associated RGF equation explicitly
involves the running cut-off $\Lambda$, and the notion of fixed point
loses its meaning here.

Therefore, we now turn to $\phi(\Lambda)$, 
for which a simple RGF equation arises, and which solution is given by:
\begin{equation}
\label{RGFsolution}
\tan\phi=(\Lambda_{0}/\Lambda)^{\alpha}\tan\phi_0
\end{equation}
where $\alpha=U_{0}/(2\pi\hbar v_F)$.
The corresponding transmission coefficient $T(\Lambda)$ on a given impurity is:
\begin{equation}
\label{transmissionvsT}
T(\Lambda)=\cos^{2}(\phi(\Lambda))=\frac{T_{0}(\Lambda/\Lambda_{0})^{2\alpha}}
{R_{0}+T_{0}(\Lambda/\Lambda_{0})^{2\alpha}}
\end{equation}
where $T_{0}$ is the transmission coefficient for a single impurity in absence
of interaction, and $R_{0}=1-T_{0}$. This result agrees with the expression
obtained for a single impurity~\cite{Yue94} in the absence of spin backscattering,
namely when $\tilde{U}(2k_{F})=0$. Again, this approach assumes small
electron-electron interactions. In the case of strong interactions, where
$K$ is no longer close to $1$, the bosonization method shows that for the
single impurity problem, $\alpha$ should be replaced by $(1-K)/2$~\cite{Kane92b}.
These two expressions for the exponent coincide at small $U_{0}$ if terms
of order $U_{0}^{2}$ or higher are neglected.

For a commensurate system ($k_{F}a=\pi n$, $n$ integer), the 
non-interacting ground-state is already gapped, so we expect
a true insulator as well in the presence of repulsive interactions.
The difference between a traditional band insulator and the one
obtained here in the presence of interactions is the
non-trivial energy dependence of the impurity scattering matrix
and the corresponding behavior of the Landauer conductance, 
Eq.~(\ref{transmissionvsT}). For an incommensurate system,
we have a partially filled band crossing the Fermi level in the
absence of interaction. Since our renormalization procedure
assumed a gradual elimination of fully occupied bands, it has
to break down after the last of those bands has been integrated out.
Treating the remaining partially occupied band in a heuristic
way, we simply assume that it corresponds to a strongly
renormalized Luttinger liquid, whose effective Fermi velocity
$v_{F}^{*}$ is much reduced compared to the Fermi velocity
$v_{F}=\hbar k_{F}/m$ of an uniform non-interacting gas with the
same density. More precisely, we have:
\[v_{F}^{*} \simeq v_{F}\cos \phi_{S}\sin(k'_{F}a)\]
where $\phi_{S}\simeq \pi/2$ is the value of $\phi$
when the renormalization procedure stops, which corresponds
to:
\[\frac{\Lambda_{0}}{\Lambda}=\frac{k_{F}a}{\pi}\]
Using Eq.~(\ref{RGFsolution}), we get:
\be
v_{F}^{*}=v_{F} \left(\frac{\pi}{k_{F}a}\right)^{\alpha}
\cot \phi_{0} \sin(k'_{F}a)
\ee

\section{Generalization of RG procedure to a large class of lattices}
\label{generallat}

In the previous section we introduced the main ideas we used to obtain the RGF equation for a 1D lattice.
We wish now to show that renormalizability of this particular 1D system is not a simple coincidence, 
but a general property of any network (not necessarily periodic), provided the two following
assumptions hold, namely all the links have the same length, which has to be large compared
to the Fermi wave-length. 
Let us begin to follow the same procedure as in one dimension.
Suppose that we have a network of equal length wires. Any junction point is described by 
an unitary $\hat S$-matrix which dimension is equal to the number of wires joining at this node. 
For each link, stationary single electron states can  
be written as the sum of two plane waves. 
\be\label{wavefunction}
\psi(x)=A_{ij}e^{-ikx}+A_{ji}e^{ikx}
\ee
where $A_{ij}$ is the amplitude of the wave that propagates from node $j$ to node $i$,
if the $x$ coordinate is oriented from $i$ to $j$.
Solving Schr\"odinger's equation is equivalent to connect these 
various amplitudes via node scattering matrices:
\be\label{schroed}
A_{ij}=\sideset{}{^{(j)}}\sum_m e^{ika}S^{(j)}_{im}A_{jm}
\ee
Here $\sideset{}{^{(j)}}\sum\limits_m$ means that we sum over first neighbors $m$ of node $j$. 
We notice that this has indeed the form of an eigenvalue equation written in some basis. 
Following the idea of Kottos and Smilansky~\cite{Kottos97} 
we introduce a finite dimensional Hilbert space associated to the lattice links. 
Each link $ij$ is represented by two orthonormal vectors $|ij\rangle$ and $|ji\rangle$. 
The dimension of this auxiliary Hilbert space is therefore $2N_L$ ($N_L$ is 
the total number of links). 
One may rewrite Eq.(\ref{schroed}) in its vector form:
\be\label{BSoperator}
\hat T |A(k)\rangle=e^{-ika}|A(k)\rangle
\ee
where the $\hat T$ operator incorporates information about the scattering matrices of all nodes.
\be\label{T}
\hat T=\sum_j\sideset{}{^{(j)}}\sum_{i,m}|ij\rangle S_{im}^{(j)}\langle jm|\Longleftrightarrow
S_{im}^{(j)}=\langle ij|\hat T|jm\rangle
\ee
As this operator $\hat T$ is unitary and defined in a finite dimensional Hilbert space, 
it could be diagonalized as: $\hat T|\alpha\rangle=e^{-i\theta_\alpha}|\alpha\rangle$, 
where $\alpha$ takes $2N_L$ values and $\theta_\alpha$ is real. 
So we obtain families of eigenvalues for the single electron energy $E=\hbar^2k^2/2m$:
\be\label{k-theta}
ak_{\alpha,n}=\theta_\alpha+2\pi n \ge0
\ee
We emphasize that this periodic structure of the single particle spectrum  
is a special feature of constant link length networks. 
A brief discussion of the more general case is given in appendix~\ref{generalization}.
Because of this periodicity, and despite the absence of any translational
symmetry, we may still introduce a notion of energy band for such lattices.
More precisely, in this setting, an energy band corresponds to fixing
$n$ and allowing for all possible values of $\theta$.
Note that this notion of band does not exactly coincide with the more familiar
notion from the Bloch theory of translational invariant lattices.
For simple Bravais lattices, the number of states in each Bloch band is
the number of unit cells which is equal to the number of sites $N_{S}$.
If $Z$ is the coordination number, we have $ZN_{S}=2N_{L}$, so our generalized
bands contain $Z$ usual Bloch bands for a Bravais lattice. 
At this stage, we have so far a band structure equation written in operator form. 
In order to obtain renormalization flow for $\hat S$-matrix we need 
to compute first the electron-electron contribution as in Eq.~(\ref{E1simple}) 
to the single electron energy and then the variation in the unperturbed energy
due to an arbitrary $\hat S$-matrix change $\partial E_0/\partial \hat S$.

As in one dimension, the main contribution to the electronic self-energy is given by the
exchange term. Let us consider a pair of single particle eigenstates labelled by $k$ and $q$, 
where these labels should in fact be viewed as pairs $(\alpha,n)$ and $(\beta,m)$,
$m$ and $n$ being integers according to the above description of the spectrum.
State $k$ is close to the Fermi level, but state $q$ is far from it, at a distance
corresponding to the current energy cut-off $\Lambda$.
Along  a link $ij$, we denote by $(\Psi^{*}_{k}(x)\Psi_{q}(x))_{0}$
the slowly varying component of $\Psi^{*}_{k}(x)\Psi_{q}(x)$.
A simple computation shows that:
\begin{widetext}
\begin{eqnarray}
\frac{1}{L}\int_{i}^{j}|(\Psi^{*}_{k}(x)\Psi_{q}(x))_{0}|^{2}  = 
|A_{ij}(k)|^{2}|A_{ij}(q)|^{2}+|A_{ji}(k)|^{2}|A_{ji}(q)|^{2} 
+(A_{ji}^{*}(k)A_{ij}(k)A_{ji}(q)A_{ij}^{*}(q)+\mathrm{h.c.})
\frac{\sin (k-q)L}{(k-q)L}
\end{eqnarray}
The first part summed over fully completed band does not depend on energy:
\be
\sum_{q\in \text{Band}}\sum_{i,j}|A_{ij}(k)|^{2}|A_{ij}(q)|^{2}+|A_{ji}(k)|^{2}|A_{ji}(q)|^{2}=\sum_{q\in \text{Band}}\sum_{<ij>}|A_{ij}(k)|^{2}|A_{ij}(q)|^{2}&=&\nonumber\\ =\sum_{<ij>}\langle ij|k\rangle\langle k|ij\rangle\langle ij|\sum_{q\in \text{Band}}|q\rangle\langle q|ij\rangle=\sum_{<ij>}\langle k|ij\rangle\langle ij|k\rangle=\langle k|k\rangle=1
\ee
We used the fact that both $|q\rangle$ and $|ij\rangle$ form complete basis sets in our Hilbert space. 
The main expression to compute is then
\be\label{I}
I(k)=\sum_{q\in\text{Band}}\sum_{<jm>} A_{jm}^*(k)A_{mj}(k)A_{jm}(q)A_{mj}^*(q)
\sin(k-q)a
\ee
\end{widetext}
where the sum over $q$ is just a single band sum, namely $m$ is fixed, and the sum is
taken over the $2N_{L}$ values of $\beta$.
The power of this algebraic formalism is that such sum is readily performed, without
having to compute any integral. Indeed, we have: 
\be
I(k)=\sum_q\sum_{<ij>}\langle k|ij \rangle\langle ij|q \rangle \sin[(k-q)a]\langle q|ji \rangle\langle ji|k \rangle
\ee
As shown in appendix~\ref{generalization}, we may assume that the eigenvectors $|q\rangle$
are normalized to unity in the auxiliary Hilbert space attached to link amplitudes,
provided the links have the same length, much larger than the Fermi wave-length.
Therefore, we have the very useful completeness relation, that is:
\be
\sum_{q}|q\rangle e^{-iqa}\langle q|=\hat{T}
\ee
After some simple algebra, we may cast $I(k)$ into the form
\be
I(k)=e^{ika}\frac{i}{\alpha}\langle k|\sum_j\sideset{}{^{(j)}}
\sum_{l,m}|lj\rangle V^{(j)}_{lm}\langle jm | k\rangle=\nonumber\\ 
\mbox{}=e^{ika}\frac{i}{\alpha}\langle k|\hat{ \bf V}| k\rangle
\ee
where the single node operators $\hat V^{(j)}$ are defined by:  
\[\hat V^{(j)}=\hat F^{(j)}-\hat S^{(j)}\hat F^{(j)\dagger}\hat S^{(j)}\]
and the diagonal matrix $\hat{F}^{(j)}$ by:
\[F^{(j)}_{ii}=-\frac{1}{2}\alpha S^{(j)}_{ii}\]
We have introduced as before the dimensionless parameter $\alpha=U_0/(2\pi\hbar v_F)$.

To get the first order variation of the single electron energy 
under small changes in the node scattering matrix parameters 
we differentiate Eq.~(\ref{BSoperator}):
\be
(d\hat T)|k\rangle+\hat T|dk\rangle=-i(dk)ae^{-ika}|k\rangle+e^{-ika}|dk\rangle
\ee
Applying $\langle k|$ to this equation and using $\langle k|\hat T=e^{-ika}\langle k|$ we obtain:
\be
dk=\frac{i}{a}e^{ika}\frac{\langle k|d\hat T|k\rangle}{\langle k|k\rangle}
\label{kvar}
\ee
This allows us to calculate single electron energy variations due to $\hat S$-matrix changes. 
In the particular case of global phase transformation, 
the corresponding infinitesimal form reads: $d\hat T=i\hat T d\psi$. 
Clearly, the energy differential does not depend on energy any more since $dk=-d\psi/a$,
so global phase shifts simply induce a global translation on the energy spectrum.

Following the same ideas as in 1D, we generalize the 
RGF equation to any $\hat S$-matrix parametrization. Eq.(\ref{factor}) now becomes:
\be\label{factorgeneral}
\frac{\partial E_0}{\partial S}(\hat S_0,k)dS(\Lambda)=E_1(\hat S_0,k,\Lambda_0)-E_1(\hat S_0,k,\Lambda)
\ee
The left-hand side of this equation is equal to $\hbar v_Fdk$, where 
$dk$ is related to the small renormalization of $\hat{T}$ by Eq.~(\ref{kvar}). 
To evaluate right-hand side, as before, we integrate just over one band of width $2\pi$ 
for the quantity $q_{\beta,m}a$, i.e. $d\Lambda=\Lambda-\Lambda_0=-2\pi$. 
\begin{widetext}
\begin{eqnarray*}
\hbar v_F dk=E_1(\hat S_0,k,\Lambda_0)-E_1(\hat S_0,k,\Lambda)=\\=\frac{cU_0(\Lambda_0-\Lambda)}{a}-\frac{U_0}{a}\sum_{\Lambda<qa< \Lambda_0}\sum_{<jm>} A_{jm}^*(k)A_{mj}(k)A_{jm}(q)A_{mj}^*(q)\frac{\sin(k-q)a}{(k-q)a}=-\frac{cU_0d\Lambda}a-\frac{U_0}{a}\frac{1}{\Lambda}\left(-\frac{d\Lambda}{2\pi}\right)I(k)
\end{eqnarray*}
\end{widetext}
Constant $c$ includes both the Hartree term and the part of exchange term that 
does not depend on $k$, so we do not precise its value since it renormalizes 
only the global phase of $\hat S$-matrix.
 
Finally, we get the result $d\hat{T}/dl=-\hat{\bf V}$  that agrees 
completely with Lal, Rao and Sen~[\onlinecite{Lal02}], obtained for a
single node connecting an arbitrary number of semi-infinite 1D wires. 
In coordinate way of writing, it gives:
\be
\frac{d\hat S^{(j)}}{dl}=\hat S^{(j)}\hat F^{(j)\dagger} \hat S^{(j)}-\hat F^{(j)}
\label{nodeRGF}
\ee
where we just chose the usual cutoff parametrization: $\Lambda=\Lambda_0e^{-l}$.

\section{Two-dimensional square lattice}
\label{2Dlat}

\begin{figure}[bt]
\includegraphics[width=2.7in]{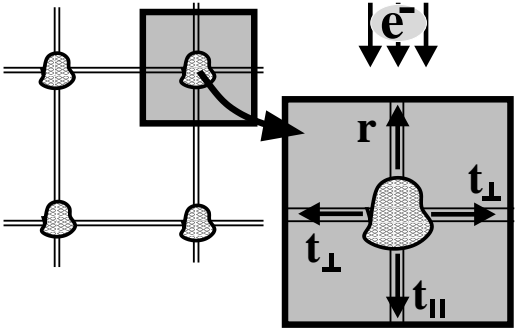}
\caption
{Two dimensional periodic grid of electron liquids with impurities. Each impurity could be represented by 3 complex parameters: $r$- reflection, $t_{\|}$- forward transmission, $t_{\bot}$- perpendicular transmission coefficients.}
\label{grid2D_xfig}
\end{figure}

We would like to illustrate the result of the previous section on one more example. 
This part could be interesting from an experimental viewpoint, since 
present nanofabrication techniques are now available
to prepare networks of quantum wires with a very small number of
transverse conduction channels etched on a two-dimensional electron gas
with high mobility, as illustrated for instance in~[\onlinecite{Naud01}].
Let us now consider an infinite regular square lattice of perfect Luttinger wires. 
These one dimensional conductors are only coupled at the lattice nodes which are described by a single $4\times 4$ scattering matrix $\hat{S}$. 
To keep a simple model, we shall restrict ourselves to the case of a single conduction 
channel in each wire, although the case of several channels would clearly be of interest, 
both on the theoretical side, and with respect to possible experimental realizations. 
As mentioned in the Introduction, we shall not take into account any energy dependence 
of the scattering matrix, although detailed studies of the Schr\"odinger equation 
for a cross of wires with a finite width have exhibited a rich pattern of 
resonances~\cite{Schult89,Berggren91}. 
The main motivation for this simplified treatment is that in a renormalization group picture, 
smooth energy dependences in the scattering matrix as a function of $E-E_{F}$ 
correspond to irrelevant operators, which should not alter drastically the way 
interactions drive the system to its low-energy fixed point. 
Labelling the four directions joining at a node as on Fig.~\ref{grid2D_xfig}, 
we shall consider a scattering matrix of the following form:
\begin{eqnarray}\label{Smatrix2D} \hat S =
\left(\begin{array}{cccc}r & \tpar & \tperp & \tperp \\
\tpar & r & \tperp & \tperp\\
\tperp & \tperp & r & \tpar\\
\tperp & \tperp & \tpar & r
\end{array}\right)
\end{eqnarray}
which corresponds to the most general form obeying time inversion and spacial
$D_{4}$ dihedral symmetry, in combination with unitarity.
The previous expressions involves three complex parameters, but as shown in
appendix~\ref{Sparameterization}, unitarity leaves only three independent real
variables. We have chosen the following parametrization:
\begin{equation}\label{r-tperp-tpar}
\begin{cases}
\begin{array}{l} 
r=e^{i\psi}(e^{2i\phi_u}+e^{2i\phi_v}-2)/4\\
\tpar=e^{i\psi}(e^{2i\phi_v}+e^{2i\phi_u}+2)/4\\
\tperp=e^{i\psi}(e^{2i\phi_v}-e^{2i\phi_u})/4\\
\end{array}
\end{cases}
\end{equation}
where $\phi_{u,v}\in[0,\pi[ \text{ and }\psi\in[0,2\pi[$\\
Note that two lines in the $(\phi_{u},\phi_{v})$ plane are specially
interesting:
\begin{equation}
\begin{cases}\begin{array}{lll}\phi_u=\pi/2 \quad &\Rightarrow \quad \tperp=\tpar\qquad &\text{(symmetric case)}
\\ \phi_u=\phi_v \quad&\Rightarrow\quad \tperp=0 \qquad&\text{(1D case)}\end{array} \end{cases}
\end{equation}

\subsection{Band Structure}
The derivation of the band structure is standard, so it is outlined in appendix~\ref{2Dbandstructure}.
This band structure is given by an implicit equation:
\begin{equation}\label{BS2Dnonsym}
x(k,\mathbf{k'})+y(k,\mathbf{k'})=
\beta\equiv\frac{2\cos\phi_v}{\sin(\phi_u-\phi_v)}\end{equation} \qquad \text{where}
\begin{eqnarray}\label{xy} 
x(k,\mathbf{k'}) & \equiv & \frac{\sin (ka+\psi)}{cos\phi_u\cos(k'_{x}a)-\cos(ka+\psi+\phi_u)}\\
y(k,\mathbf{k'}) & \equiv & \frac{\sin (ka+\psi)}{cos\phi_u\cos(k'_{y}a)-\cos(ka+\psi+\phi_u)}
\end{eqnarray}
As usual, the energy of these states is given by the free electron dispersion
$E_{0}(k)=\hbar^{2}k^{2}/(2ma^{2})$.
Here, $\mathbf{k'}$ is the two-dimensional lattice wave-vector, such that
$\Psi(\mathbf{r+R})=\exp(i\mathbf{k'.R})\Psi(\mathbf{r})$ for any $\mathbf{r}$
on the wire network and any period $\mathbf{R}$ of the square lattice.

\begin{figure}[h]
\includegraphics[width=0.95\linewidth]{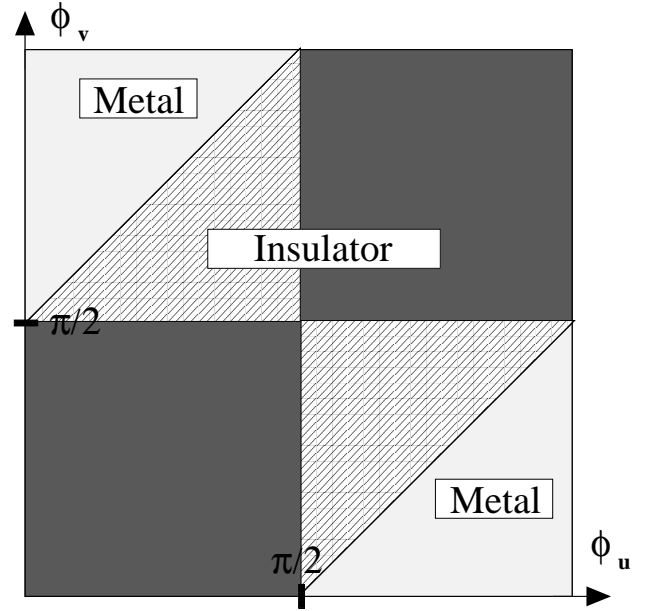}
\caption
{Phase diagram for a non-interacting electron wire square grid. 
Contrary to the one dimensional case, there are metallic states at
integer filling factors for some values of the $\hat S$-matrix.
In these regions of the phase diagram, the single electron spectrum
is gapless.}
\label{RGF2D_xfig}
\end{figure}

As we found some interesting features in the band structure of noninteracting electrons in 
a two dimensional square grid we will describe it more precisely. 
Contrary to one dimension, there are values of 
the scattering matrix, for which the single electron spectrum is no longer
gapped, and these are located on Fig.~\ref{RGF2D_xfig}.
More precisely, in the clear regions of Fig.~\ref{RGF2D_xfig}, the single
particle spectrum is gapless. In the dark regions, it is gapped, leading to
an insulator if the electronic density corresponds to filling an
{\em even integer} number of bands. Finally, in the dashed regions, we obtain
an insulator for an {\em odd integer} number of bands.

We still have a $2\pi$ periodic structure in $ka$, but the band-structure 
consists of two types of foils: 
normal and abnormal. Normal bands resemble an ordinary band of a tight-binding model of 
square lattice crystal (a sort of deformed paraboloid). 
Abnormal bands are so called for their strange curvature. To get an idea of their form 
one could imagine a square rubber foil, attach its four extremities 
and then put inside a heavy cross. For a complete description,  
we give sections of the band structure in several directions for three 
characteristic values of the scattering matrix.
\begin{figure}[bt]
\includegraphics[width=0.85\linewidth]{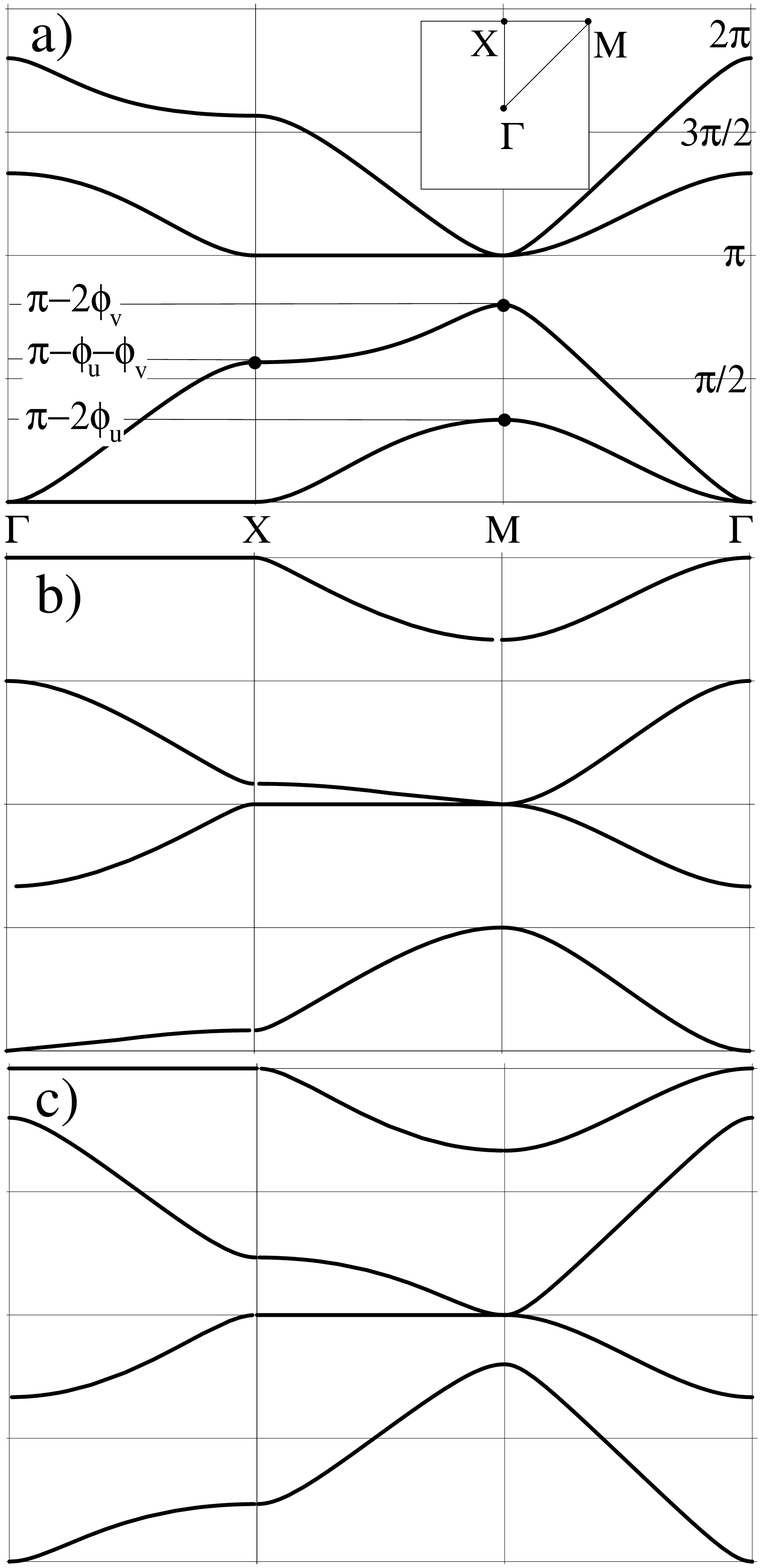}
\caption
{3 characteristic band structure pictures for different value of $\hat S$-matrix:\\
(a) Insulator, $0<\phi_u,\phi_v<\pi/2$, (dark on Fig.\ref{RGF2D_xfig})\\
(b) Insulator, $0<\phi_u<\pi/2,\pi/2<\phi_v<\pi,|\phi_u-\phi_v|<\pi/2$, (dashed on Fig.\ref{RGF2D_xfig})\\
(c) Conductor, $0<\phi_u<\pi/2,\pi/2<\phi_v<\pi,|\phi_u-\phi_v|>\pi/2$, (clear on Fig.\ref{RGF2D_xfig})\\
The band-structure is $2\pi$ periodic in $k$, and has 4 foils:2 normal and 2 abnormal. 
Some foils are described as ''abnormal'' because of their strange curvature, revealed here
by the flat part of these bands. Given the energy interval $0<k<\pi$ one could obtain 
the $\pi<k<2\pi$ interval by exchanging $\Gamma$ and $M$ points.}
\label{BStr_joint}
\end{figure}
Because of some important symmetries, we may restrict the domain of variation of 
\{$\phi_u,\phi_v$\}, and still get all the possible different physical pictures:
\begin{enumerate}
\item $k(\phi_u,\phi_v)=k(\phi_v,\phi_u)$, 
\item $k(\pi-\phi_v,\pi-\phi_u)=-k(\phi_u,\phi_v)$.
\end{enumerate} 
These may be easily seen from form II of the dispersion relation, given in appendix B. 
Both of them are reflection symmetries.  
Given the band-structure for $k \in [0,\pi]$ and using 
the following symmetry: $k(k^{\prime}_x,k^{\prime}_y)+\pi=k(k^{\prime}_x+\pi,k^{\prime}_y+\pi)$, 
we easily expand it to the full interval $k\in[0,2\pi]$ by 
replotting the same band originating from point $M$ instead of $\Gamma$ 
(see Fig.~\ref{BStr_joint}).

\subsection{RGF equation for a two dimensional grid}
 
\begin{figure}[bt]
\includegraphics[width=0.95\linewidth]{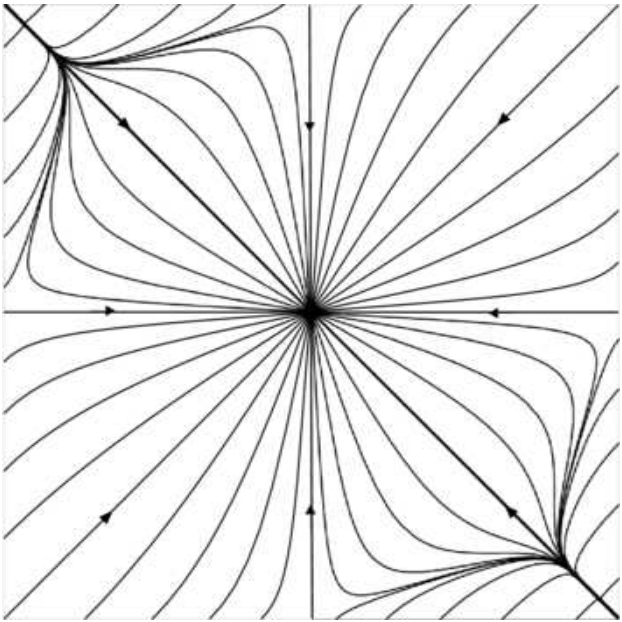}
\caption
{RGF for 2D square grid of wire. The only attractor is in the picture center for $\{\phi_u,\phi_v\}=(\pi/2,\pi/2)$.}
\label{RGFexact_xfig}
\end{figure}

Following the same procedure as in the one dimensional case,
we first calculate Hartree and exchange contributions to single electron energy 
and then establish the equivalent of Eq.(\ref{factor}) or Eq.~(\ref{factorgeneral}) 
for two dimensions and finally get the RGF equation. The main difference with the 1D case 
is that we have now three real parameters for the $\hat S$-matrix 
and electron-electron interactions should be evaluated 
along two perpendicular threads that form our grid.
The condition to satisfy now reads:
\begin{eqnarray} \label{factor3} \frac{\partial E_0}{\partial \psi}\psibar(\Lambda)+\frac{\partial
E_0}{\partial \phi_u}\phibar_u(\Lambda)+ \frac{\partial E_0}{\partial
\phi_v}\phibar_v(\Lambda)=\nonumber\\ \mbox{}=\mathcal E_1(\hat S_0,\mathbf{k'},\Lambda_0)-\mathcal E_1(\hat S_0,
\mathbf{k'},\Lambda)
\end{eqnarray}
As was proven in the previous section all networks with links of equal length 
are renormalizable i.e. there is a set of functions $\phibar_u,\phibar_v$ and $\psibar$ 
depending only on $\Lambda$. Indeed the decomposition of 
the r.h.s. of Eq.~(\ref{factor3}) on a basis of three functions depending on $\mathbf{k'}$ 
is possible.
The corresponding renormalization group flow equations are:  
\begin{eqnarray}
\frac{d\phi_u}{dl} & = & \frac{\alpha}{8}(\sin2\phi_v+3\sin2\phi_u+\sin2(\phi_v-\phi_u))\\
\frac{d\phi_v}{dl} & = & \frac{\alpha}{8}(\sin2\phi_u+3\sin2\phi_v+\sin2(\phi_u-\phi_v))
\end{eqnarray}
where $\alpha$ is defined in Eq.~(\ref{BS2Dnonsym},\ref{xy}).
The only fixed points are $\phi_{u,v}=0,\pi/2$, among which
there is only one attractor for $\{\phi_u,\phi_v\}=\{\pi/2,\pi/2\}$. 
The global behavior of this flow is illustrated on Fig.~\ref{RGFexact_xfig}.
These properties of the RGF for a single node connecting four
semi-infinite wires have already been described by Lal et al.~\cite{Lal02}
and Das et al.~\cite{Das03}. As for the one dimensional example
of section~\ref{1Dlat} above, the new feature associated to a regular lattice
is the presence of commensurability effects. We have to stop the
renormalization procedure when all the completely filled bands
have been eliminated. From Fig.~\ref{BStr_joint}, we expect
to obtain one or two partially filled bands crossing the Fermi level.
These bands are only very weakly dispersive, since the effective
$\hat{S}$ matrix for the nodes is then very close to its value
at the vanishing transmission fixed point. Suppose now that this
fixed point is approached from the dark regions of the phase-diagram
shown on Fig.~\ref{RGF2D_xfig}. If the filling factor 
corresponds to an {\em even integer}, the Fermi level lies in a gap
of the renormalized band structure. Therefore, we may eliminate
the remaining pair of filled bands, and the system is an insulator.
Similarly, a true insulator is obtained for an {\em odd integer}
filling factor, in the case where the $(\pi/2,\pi/2)$ fixed point
is approached from the dashed regions in Fig.~\ref{RGF2D_xfig}.
Experimentally, one expects transitions between
these commensurate insulators and strongly renormalized
``heavy electron'' metals at generic filling factors
if the electronic density is controlled by a 
uniform external gate voltage.

Another interesting feature of this geometry is the fact that
the flow may induce metal-insulator transitions for some
commensurate filling factors at a {\em finite} energy scale.
Indeed, for initial parameters lying in the clear regions
of Fig.~\ref{RGF2D_xfig}, corresponding to a gapless single
electron spectrum, Fig.~\ref{RGFexact_xfig} shows that
the system always reaches either the dashed or dark regions
in a finite RG time. Experimentally, these RG flows
may be visualized by gradually lowering the temperature,
since at least qualitatively, the energy scale set by temperature
plays the role of the moving cut-off $\Lambda$. 

\section{Conclusion}

In this paper, we have studied a particular class
of networks of Luttinger liquids, with nodes connected by links
of a constant length. In the limit of long links, compared
to the Fermi wave-length, we studied the evolution of
the scattering matrix at the nodes, as the typical energy scale
for the occupied states contributing to Friedel oscillations
is getting closer to the Fermi level. 
The corresponding renormalization group flow turns 
to be identical to the one already found for a single node
coupled to several semi-infinite 1D Luttinger liquids~\cite{Lal02}.
This result is physically reasonable, since we have considered
the limit of long links. However, we emphasize that these renormalization
effects come from quasiparticle scattering on Friedel oscillations
induced by the nodes, which are a rather complicated function of the
lattice geometry. For instance, even in the limit of very long links,
the amplitudes $A_{ij}$ which determine the value of energy eigenfunctions
along the links are obtained from a $2N_{L}\times 2N_{L}$ eigenvalue
problem whose solution has a strongly non-local character.

The main difference between a regular lattice and a simple
node coupled to infinite wires is that in the former
case, we have to stop the renormalization procedure when the
last occupied band has been integrated out. So instead of having
completely disconnected wires in the low-energy limit, 
we expect in general a strongly renormalized conducting system
with an effective Fermi velocity much reduced in comparison
to a non-interacting system with the same density.
These effects should be visible as a power-law behavior of
the network conductance as a function of temperature. 
Insulating ground-states are expected when the electronic density 
corresponds to filling some integer numbers of bands.  

Of course, this work leaves many open questions.
It would be interesting to generalize the present
renormalization approach to lattices containing links
with several different lengths. In such situations,
the spectrum no longer exhibits a simple periodic structure,
and some signatures of quantum chaos, already manifested in
the single particle density of states~\cite{Kottos97},
may also appear in the temperature dependence of the conductivity
of an interacting system. Another open question is the influence
of an external magnetic field, which also drastically
modifies the single-particle spectrum. Finally, the limit of
strong electron-electron interaction deserves further investigation,
and in particular the possibility to develop some new metal-insulator
transitions for non-integer but rational filling factors,
generalizing the notion of a Wigner crystal. Such insulating 
states would naturally be pinned by the nodes of the lattice.

\begin{acknowledgments} 
We would like to thank J. Dufouleur, G. Faini, D. Mailly,
C. Naud, and J. Vidal for interesting discussions on various
aspects of conducting wire networks.
\end{acknowledgments}

\appendix

\section{Parameterization of the $\hat{S}$ matrix}
\label{Sparameterization}

In this appendix we will show that time-inversion, spacial $D_4$ dihedral symmetry 
combined with unitarity imply a parameterization of scattering matrix 
in terms of three real variables.
For two-dimensional square lattice, the most general form of $\hat S$-matrix 
is given by a 4x4 matrix:
\begin{eqnarray} 
\left(\begin{array}{c}A^{\prime}\\B^{\prime}\\C^{\prime}\\D^{\prime}\end{array}\right)=
\left(\begin{array}{cccc}r_A & t_{BA}& t_{CA}&t_{DA} \\
t_{AB}& r_{B}& t_{CB}& t_{DB}\\
t_{AC}& t_{BC}& r_C& t_{DC}\\
t_{AD}& t_{BD}& t_{CD}& r_D
\end{array}\right)
\left(\begin{array}{c}A\\B\\C\\D\end{array}\right)
\end{eqnarray}
\begin{figure}[bt]
\includegraphics[width=0.4\linewidth]{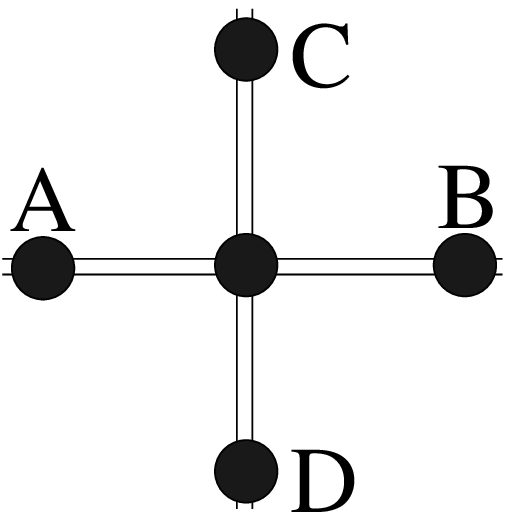}
\end{figure}
where $A,B,C,D$ are the coefficients of incoming plane waves. 
Primed values denote coefficients of outgoing waves. 
At this stage, one has 16 complex parameters for this $\hat S$-matrix.

Unitarity condition ($\hat S^\dagger \hat S = I$) combined with 
time-inversion symmetry ($\hat S^{-1}=\hat S^*$) gives $\hat S^t=\hat S$ 
(notice that $\hat S^t$ is the transposed matrix, not the conjugate).
It leaves 10 complex parameters. Using four reflections of two types 
(1) $A\leftrightarrow B,$ and (2) $A\leftrightarrow C$, $B\leftrightarrow D$ 
that generate the dihedral symmetry group $D_4$ consequently reduces this number to 
three complex variables. We obtain the $\hat S$-matrix in the form~(\ref{Smatrix2D}). 
Unitarity allows finally to express the scattering matrix with only 3 real parameters:
\begin{equation*}
  \begin{cases}
     |r|^2+2|t_{\bot}|^2+|t_{\|}|^2 = 1 \\
     r t_\bot ^* + r^* t_\bot + \tperp^* \tpar +\tpar^* \tperp = 0 \\
     r \tpar^* + r^* \tpar + 2|\tperp|^2 = 0 \\
   \end{cases}
\end{equation*}
Subtracting the third equation from the first one allows us to define 
a first real parameter $\psi$:
\[|r-\tpar|=1\Rightarrow\ r=\tpar-e^{i\psi}\]
There remains two independent equations:
\begin{equation*}
  \begin{cases}
     \hspace{-1.3cm}\text{I }\hspace{1.1cm} 2(|\tpar|^2+|\tperp|^2) = \tpar e^{-i\psi}+\tpar^* e^{i\psi} \\
     \hspace{-1.3cm}\text{II}\hspace{1.1cm} 2(\tperp^*\tpar+\tperp\tpar^*) = \tperp e^{-i\psi}+\tperp^* e^{i\psi}\\
   \end{cases}
\end{equation*}
\begin{equation*}
  \begin{cases}
     \hspace{-1.3cm}\text{I--II}\hspace{0.7cm} 2|\tpar+\tperp|^2 = 2\text{Re}[(\tpar+\tperp) e^{-i\psi}] \\
     \hspace{-1.3cm}\text{I+II}\hspace{0.65cm} 2|\tpar-\tperp|^2 = 2\text{Re}[(\tpar-\tperp) e^{-i\psi}] \\
   \end{cases}
\end{equation*}
Two more real parameters are needed to complete the parametrization:
\begin{equation*}
  \begin{cases}
     \tpar-\tperp=\cos{\phi_u}e^{i(\phi_u+\psi)}\\
     \tpar+\tperp=\cos{\phi_v}e^{i(\phi_v+\psi)}
   \end{cases}
\end{equation*}
Expressions of transmission and reflection coefficients as functions of these 
three real parameters are given in the main text, see  Eq.~(\ref{r-tperp-tpar}).
We remark an interesting fact: in the case of perfect transmission ($r=0$), 
only the separate thread solution ($|\tpar|=1,\tperp=0$) is possible.

\section{Band structure for a square lattice of wires}
\label{2Dbandstructure}
\begin{figure}[bt]
\includegraphics[width=2.7in]{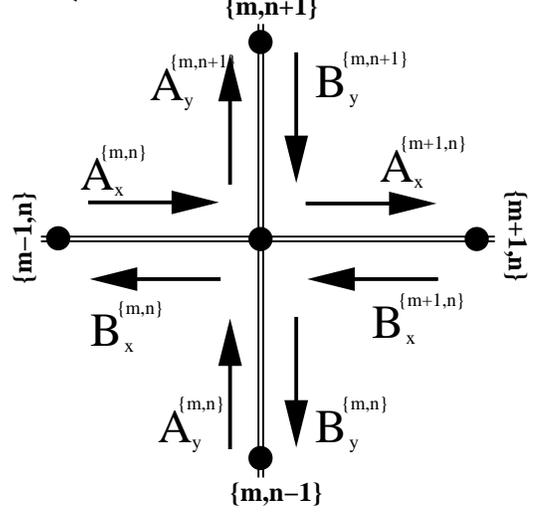}
\caption
{In 2D case, each node is indexed by a pair of numbers \{m,n\}. Incoming and outgoing 
plane waves are connected by the $4 \times 4$ scattering matrix.}
\label{single_node2D_xfig}
\end{figure}

In this appendix we derive the band-structure for a  square lattice of wires 
of non-interacting electrons. As in the one dimensional case, the wave-function away from impurities (i.e. nodes here) could be written as combination of plane waves:
\begin{equation}
\psi_k(x)=A_i^{m,n}e^{ikx}+B_i^{m,n}e^{-ikx}
\end{equation}
where $i=\{x,y\}$ and the coefficients
$\{A_x^{m,n},A_y^{m,n},B_x^{m,n},B_y^{m,n}\}_{m,n}$ 
are defined on Fig.~\ref{single_node2D_xfig}.
By definition of the scattering matrix:
\begin{widetext}
\begin{equation}
\left(\begin{array}{l}
B_x^{m,n}\\ A_x^{m+1,n} \\A_y^{m,n+1}\\B_y^{m,n}\end{array}\right) = \hat S
\left(\begin{array}{l} A_x^{m,n} \\
B_x^{m+1,n}\\B_y^{m,n+1}\\A_y^{m,n}\end{array}\right)=\left(\begin{array}{cccc}r & \tpar & \tperp & \tperp \\
\tpar & r & \tperp & \tperp\\
\tperp & \tperp & r & \tpar\\
\tperp & \tperp & \tpar & r
\end{array}\right)
\left(\begin{array}{l} A_x^{m,n} \\
B_x^{m+1,n}\\B_y^{m,n+1}\\A_y^{m,n}\end{array}\right)
\end{equation}
Bloch periodicity condition for the wave-function implies:
\begin{equation}
\begin{cases}\begin{array}{lcl} A_x^{n,m}&=& e^{i(k^{\prime}_x-k)na}e^{ik^{\prime}_yma}A_x^{0,0}\\
B_x^{n,m}&=& e^{i(k^{\prime}_x+k)na}e^{ik^{\prime}_yma}B_x^{0,0}\\
A_y^{n,m}&=& e^{i(k^{\prime}_y-k)ma}e^{ik^{\prime}_xna}A_y^{0,0}\\
A_x^{n,m}&=& e^{i(k^{\prime}_y+k)ma}e^{ik^{\prime}_yna}B_y^{0,0}\\
\end{array}\end{cases}
\end{equation}
Using the last two expressions, we obtain the secular equation:
\begin{eqnarray}
\left|\begin{array}{cccc}
\tpar e^{i(k-k^{\prime}_x)a}-1& re^{2ika}& \tperp e^{i(k-k^{\prime}_x)a}&\tperp e^{i(k^{\prime}_y-k^{\prime}_x+2k)a}
\\r&\tpar e^{i(k^{\prime}_x+k)a}-1&\tperp&\tperp e^{i(k^{\prime}_y+k)a}
\\\tperp e^{i(k-k^{\prime}_y)a}&\tperp e^{i(k^{\prime}_x-k^{\prime}_y+2k)a}&\tpar e^{i(k-k^{\prime}_y)a}-1&re^{2ika}
\\\tperp&\tperp e^{i(k^{\prime}_x+k)a}&r&\tpar e^{i(k^{\prime}_y+k)a}-1
\end{array}\right|=0
\end{eqnarray}
Replacing scattering matrix elements by their parametrization~(\ref{r-tperp-tpar}),
we get the implicit band-structure equation given in the main text~(\ref{xy}). 
Here we propose two more different ways to write the same dispersion relation,
where the combinations $ka+\psi$, $k'_{x}a$ and $k'_{y}a$
have been replaced respectively by the simpler notations
$k$, $k'_{x}$ and $k'_{y}$:
\[\text{I}\hspace*{0.5cm}
 \cos(k+\phi_u)\cos(k+\phi_v)+\cos k^{\prime}_x\cos k^{\prime}_y\cos\phi_u\cos\phi_v
 =\frac{1}{2}(\cos k^{\prime}_x+\cos k^{\prime}_y)(\cos\phi_u\cos(k+\phi_v)+\cos\phi_v\cos(k+\phi_u))
\]
\[\hspace{-4.5cm}\text{II}\hspace*{4.1cm}\frac{\cos(k+\phi_u)-\cos\phi_u\cos k^{\prime}_x}{\cos(k+\phi_u)-\cos\phi_u\cos k^{\prime}_y}=-
\frac{\cos(k+\phi_v)-\cos\phi_v\cos k^{\prime}_x}{\cos(k+\phi_v)-\cos\phi_v\cos k^{\prime}_y}\]
The first form is useful to identify symmetries of band-structure. 
The second form is useful to derive RGF equations directly without
using the formalism developed in section~\ref{generallat}. 
We remark that in the 1D case ($\tperp=0$), and in the 2D symmetric case ($\tperp=\tpar$) 
the band-structure equations are the same, namely:
$\cos(ka+\psi+\phi)=\cos\phi\cos(k^{\prime}a)$
\end{widetext}

\section{Any lattice generalization}
\label{generalization}

In this part we will discuss particular points met 
in section~\ref{generallat} of this article. 
First of all we could obtain the dispersion relation for any network 
i.e. when the wires lengths are not necessarily equal. 
In that case Eq.~(\ref{schroed}) is modified into:
\be\label{schroedfull}
A_{ij}=\sideset{}{^{(j)}}\sum_m e^{ikL_{ij}/2}S^{(j)}_{im}e^{ikL_{jm}/2}A_{jm}
\ee
We choose the origin of coordinates needed to define the amplitudes
$A_{ij}$ at the centers of each link. 
This formula means that the amplitude of the wave going from node $j$ to node $i$ 
is the sum of amplitudes coming from all neighbors $m$ of node $j$, 
multiplied by phase factors $\exp(ikL_{jm}/2)$ due to propagation from the middle of link 
$\langle jm\rangle$ to the node $j$, then scattered on node $j$ 
with probability amplitude $S_{im}^{(j)}$ and finally reaching the middle of link 
$\langle ij\rangle$ with a new phase factor $\exp(ikL_{ij}/2)$.
We will now write the same equation in vector form. 
The expression will be more transparent and this 
permits us to express the secular equation for energy eigenvalues $k$
in a compact form. 
If we fix the energy of the system then the stationary states are completely 
determined by $2N_L$ amplitudes, where $N_L$ is the number of links. 
The factor 2 arises since each wave can propagate in two opposite directions on each link. 
So the set of amplitudes $\{A_{ij}\}$ could be presented as a vector $|A\rangle$ in 
a $2N_L$-dimensional Hilbert space. 
We choose the orthonormal basis associated with network links 
$\langle mn|ij\rangle=\delta_{mi}\delta_{nj}$. 
Each link is represented by two basis vector $|ij\rangle$ and $|ji\rangle$, 
this orientation difference should be taken into account 
in various summations over first neighbors. 
We define the vector $|A\rangle=\sum\limits_{<ij>}A_{ij}|ij\rangle$ and 
the length operator $\hat L=\sum\limits_{<ij>}L_{ij}|ij\rangle\langle ij|$.
The vector form of Eq.~(\ref{schroedfull}) reads:
\be
|A(k)\rangle=e^{ik\hat L/2}\hat T e^{ik\hat L/2}|A(k)\rangle
\ee
The possible values of $k$ are given by:
\be
\det(e^{-ik\hat L}-\hat T )=0
\ee 
One remarks that the periodicity of the spectrum~(\ref{k-theta}) 
is lost in the general case, unless there exists a $\Delta k$ 
such that $\exp(i\Delta k\hat L)=1$. 
Let us remind that this periodicity of the spectrum  
allowed us to evaluate the integral $I(k)$ in Eq~(\ref{I}): 
the contribution of each band was the same and we replaced the sum over 
any filled band just by sum over all the eigen-vectors of operator $\hat T$. 
 
The second point to be clarified  is the spectral decomposition 
$\sum_q |q\rangle e^{-iqa}\langle q|=\hat T$. It holds only if $|q\rangle$ 
vectors form an orthonormal basis. 
Orthogonality is clear as $|q\rangle$ is an eigen-vector of an unitary operator.
\be
\langle q|\hat T| k\rangle=e^{-iqa}\langle q| k\rangle=e^{-ika}\langle q| k\rangle
\ee   
Let us now evaluate the vector norm in the $|ij\rangle$ basis:
\be
\langle q|q\rangle=\sum_{<ij>}\langle q|ij\rangle\langle ij|q\rangle=\sum_{<ij>} |A_{ij}(q)|^2
\ee
But we know that the norm of wavefunction~(\ref{wavefunction}) 
in the physical Hilbert space should be equal to unity.
\begin{widetext}
\be
\int_{\text{network}}|\psi(x)|^2dx=1=a\sum_{<ij>}\left(|A_{ij}(q)|^2
+A_{ij}(q)A_{ji}^{*}(q)\frac{\sin(ka)}{ka}\right)
\ee
\end{widetext}
So if we demand $\langle q|q\rangle=1$ we are doing an approximation neglecting the term 
proportional to $\sin(ka)/ka$. This approximation is legitimate in our case, 
as we consider systems where the typical number ef electrons along each link between two nodes
is large. Clearly, it will break down for links of the order of the Fermi
wave-length.
Supposing that $\langle q|q\rangle=1$ is equivalent to identify 
the norm in the (infinite dimensional) physical Hilbert space, 
with the norm associated to the orthonormal basis $|ij\rangle$ in the 
($2N_{L}$ dimensional) auxiliary Hilbert space.  
The fact that our equations are not depending explicitly on the network scale parameter $a$ 
is closely related to this approximation. 
So if one were to estimate finite size corrections 
to the RGF equation, one should take the physical normalization of 
the $|q\rangle$-basis into account.
Such corrections would likely produce RGF equations where the
nodes on the lattice are no longer renormalized independently
of each other, by contrast to what we obtained in section~\ref{generallat},
see Eq.~(\ref{nodeRGF}).


\begin{thebibliography}{99}

\bibitem{Timp91} G. Timp, R. E. Behringer, E. H. Westerwick, and J. E.
Cunningham, in {\em Quantum Coherence in Mesoscopic systems}, edited
by B. Kramer (Plenum, New-York, 1991).

\bibitem{Washburn92} S. Washburn, and R. A. Webb, Rep. Prog. Phys.
{\bf 55}, 1311, (1992)

\bibitem{Timp87} G. Timp, A. M. Chang, J. E. Cunningham, T. Y. Chang,
P. Mankiewich, R. Behringer, and R. E. Howard, Phys. Rev. Lett. {\bf 58},
2814, (1987)

\bibitem{Pedersen00} S. Pedersen, A. E. Hansen, A. Kristensen,
C. B. Sorensen, and P. E. Lindelof, Phys. Rev. {\bf B 61}, 5457, (2000)

\bibitem{Mailly93} D. Mailly, C. Chapelier, and A. Benoit,
Phys. Rev. Lett. {\bf 70}, 2020, (1993).

\bibitem{Rabaud01} W. Rabaud, L. Saminadayar, D. Mailly, K. Hasselbach,
A. Beno\^{\i}t, and B. Etienne, Phys. Rev. Lett. {\bf 86}, 3124, (2001).

\bibitem{Roukes87} M. L. Roukes, A. Scherer, S. J. Allen Jr., H. G. Craighead,
R. M. Ruthen, E. D. Beebe, and J. P. Harbison, Phys. Rev. Lett. {\bf 59}, 3011, (1987) 

\bibitem{Ford88} C. J. B. Ford, T. J. Thornton, R. Newbury, M. Pepper, H. Ahmed,
D. C. Peacock, D. A. Ritchie, J. E. F. Frost, and G. A. C. Jones, Phys. Rev.
{\bf B 38}, 8518, (1988).

\bibitem{Ravenhall89}  D. G. Ravenhall, H. W. Wyld, and R. L. Schult, Phys. Rev. Lett.
{\bf 62}, 1780, (1989).

\bibitem{Kirczenow89} G. Kirczenow, Phys. Rev. Lett. {\bf 62}, 2993, (1989).

\bibitem{Baranger89} H. U. Baranger, and A. D. Stone, Phys. Rev. Lett. {\bf 63}, 414, (1989). 

\bibitem{Beenakker89} C. W. J. Beenakker, and H. van Houten, Phys. Rev. Lett. {\bf 63},
1857, (1989).

\bibitem{Ford89} C. J. B. Ford, S. Washburn, M. B\"uttiker, C. M. Knoedler,
and J. M. Hong, Phys. Rev. Lett. {\bf 62}, 2724, (1989).

\bibitem{Naud01} C. Naud, G. Faini, and D. Mailly, Phys. Rev. Lett. {\bf 86}, 5104, (2001).

\bibitem{Vidal98} J. Vidal, R. Mosseri, and B. Dou\c{c}ot, Phys. Rev. Lett. 
{\bf 81}, 5888, (1998)

\bibitem{Vidal00} J. Vidal, G. Montambaux, and B. Dou\c{c}ot, Phys. Rev. 
{\bf B 62} R16294 (2000)

\bibitem{Grayson98} M. Grayson, D. C. Tsui, L. N. Pfeiffer, K. W. West,
and A. M. Chang, Phys. Rev. Lett. {\bf 80}, 1062, (1998).

\bibitem{Wen90} X. G. Wen, Phys. Rev. {\bf B 41}, 12838, (1990).

\bibitem{Kane94} C. L. Kane, and M. P. A. Fisher, Phys. Rev. Lett. {\bf 72}, 724,
(1994).

\bibitem{Picciotto97} R. de-Picciotto, M. Reznikov, M. Heiblum, V. Umansky,
G. Bunin, and D. Mahalu, Nature, {\bf 389}, 162, (1997).

\bibitem{Saminadayar97} L. Saminadayar, D. C. Glattli, Y. Jin, and B. Etienne,
Phys. Rev. Lett. {\bf 79}, 2526, (1997).

\bibitem{Bockrath99}M. Bockrath, D. H. Cobden, J. Lu, A. G. Rinzler,
R. E. Smalley, L. Balents, P. L. McEuen, Nature {\bf 397}, 598, (1999).

\bibitem{Yao99} Z. Yao, H. W. C. Postma, L. Balents, C. Dekker,
Nature {\bf 402}, 273, (1999).

\bibitem{Egger97} R. Egger, and A. O. Gogolin, Phys. Rev. Lett.
{\bf 79}, 5082, (1997).

\bibitem{Kane97} C. Kane, L. Balents, and M. P. A. Fisher,
{\bf 79}, 5086, (1997).

\bibitem{Papadopoulos00} C. Papadopoulos, A. Rakitin, J. Li, A. S. Vedeneev,
and J. M. Xu, Phys. Rev. Lett. {\bf 85}, 3476, (2000).

\bibitem{Kim01} J. Kim, K. Kang, J.-O Lee, H.-H. Yoo,
J.-R. Kim, J. W. Park, H. M. So, and J.-J. Kim,
J. Phys. Soc. Jpn. {\bf 70}, 1464, (2001)

\bibitem{Terrones02} M. Terrones, F. Banhart, N. Grobert, J.-C. Charlier,
H. Terrones, and P. M. Ajayan, Phys. Rev. Lett. {\bf 89}, 75505, (2002).

\bibitem{Gao03} B. Gao, A. Komnik, R. Egger, D. C. Glattli, and
A. Bachtold, e-preprint, arXiv:cond-mat/0311645.

\bibitem{Komnik98} A. Komnik and R. Egger, Phys. Rev. Lett. {\bf 80}, 2881, (1998).

\bibitem{Nayak99} C. Nayak, M. P. A. Fisher, A. W. W. Ludwig, and H. H. Lin,
Phys. Rev. {\bf B 59}, 15694, (1999).

\bibitem{Lal02} S. Lal, S. Rao, and D. Sen, Phys. Rev. {\bf B 66}, 165327, (2002).

\bibitem{Chen02} S. Chen, B. Trauzettel, and R. Egger, 
Phys. Rev. Lett. {\bf 89}, 226404, (2002).

\bibitem{Chamon03} C. Chamon, M. Oshikawa, and I. Affleck, 
Phys. Rev. Lett. {\bf 91}, 206403, (2003).

\bibitem{Das03} S. Das, S. Rao, and D. Sen,  e-preprint, arXiv:cond-mat/0311563.

\bibitem{Schult90} R. L. Schult, H. W. Wyld, and D. G. Ravenhall, Phys. Rev. {\bf B 41}, 12760,
(1990).

\bibitem{Deo94} P. Singha Deo, and A. M. Jayannavar, Phys. Rev. {\bf B 50}, 11629, (1994).

\bibitem{Uryu96} S. Uryu, and T. Ando, Phys. Rev. {\bf B 53}, 13613, (1996).

\bibitem{Kane92a} C. L. Kane, and M. P. A. Fisher, Phys. Rev. Lett. {\bf 68}, 1220, (1992).

\bibitem{Kane92b} C. L. Kane, and M. P. A. Fisher, Phys. Rev. {\bf B 46}, 15233, (1992).

\bibitem{Yue94} D. Yue, L. I. Glazman, and K. A. Matveev, Phys. Rev. {\bf 49}, 1966,
(1994)

\bibitem{Meden02a} V. Meden, W. Metzner, U. Schollw\"ock, and K. Sch\"onhammer,
Phys. Rev. {\bf B 65}, 45318, (2002).

\bibitem{Meden02b} V. Meden, W. Metzner, U. Schollw\"ock, and K. Sch\"onhammer,
Journ. Low. Temp. Phys. {\bf 126}, 1147, (2002).

\bibitem{Meden03} V. Meden, S. Andergassen, W. Metzner, U. Schollw\"ock,
and K. Sch\"onhammer, Europhys. Lett. {\bf 64}, 769, (2003).

\bibitem{Anderson70} P. W. Anderson, J. Phys. {\bf C 3}, 2346, (1970).

\bibitem{Kottos97} T. Kottos, and U. Smilansky, Phys. Rev. Lett. {\bf 79}, 4794, (1997).

\bibitem{Schult89} R. L. Schult, D. G. Ravenhall, and H. W. Wyld, Phys. Rev. {\bf B 39}, 5476,
(1989).

\bibitem{Berggren91} K.-F. Berggren, and Z.-L. Ji, Phys. Rev. {\bf B 43}, 4760, (1991).






\end{thebibliography}
\end{document}